# Nematic topological superconducting phase in Nb-doped Bi$_2$Se$_3$

Junying Shen[1], Wen-Yu He[1], Noah Fan Qi Yuan[1], Zengle Huang[1], Chang-woo Cho[1], Seng Huat Lee[2], Yew San Hor[2], Kam Tuen Law[1] and Rolf Lortz[1]

[1]Department of Physics, The Hong Kong University of Science and Technology, Clear Water Bay, Kowloon, Hong Kong

[2] Department of Physics, Missouri University of Science and Technology, Rolla, MO 65409

Correspondence: Rolf Lortz (lortz@ust.hk)

A nematic topological superconductor has an order parameter symmetry, which spontaneously breaks the crystalline symmetry in its superconducting state. This state can be observed, for example, by thermodynamic or upper critical field experiments in which a magnetic field is rotated with respect to the crystalline axes. The corresponding physical quantity then directly reflects the symmetry of the order parameter. We present a study on the superconducting upper critical field of the Nb-doped topological insulator Nb$_x$Bi$_2$Se$_3$ for various magnetic field orientations parallel and perpendicular to the basal plane of the Bi$_2$Se$_3$ layers. The data were obtained by two complementary experimental techniques, magnetoresistance and DC magnetization, on three different single crystalline samples of the same batch. Both methods and all samples show with perfect agreement that the in-plane upper critical fields clearly demonstrate a two-fold symmetry that breaks the three-fold crystal symmetry. The two-fold symmetry is also found in the absolute value of the magnetization of the initial zero-field-cooled branch of the hysteresis loop and in the value of the thermodynamic contribution above the irreversibility field, but also in the irreversible properties such as the value of the characteristic irreversibility field and in the width of the hysteresis loop. This provides strong experimental evidence that Nb-doped Bi$_2$Se$_3$ is a nematic topological superconductor similar to the Cu- and Sr-doped Bi$_2$Se$_3$.

## INTRODUCTION

The search for topological superconductors, which support Majorana bound states as low energy excitations, is one of the central topics of current research[1-5]. This is because of the fact that Majorana bound states are non-Abelian particles that have potential applications in engineering in the form of topologically protected qubits for quantum computation[6-8]. A promising approach to the realization of topological superconductors is to use *s*-wave superconductors to induce superconducting pairing on topological insulators[9] or systems with strong Rashba spin-orbit coupling[10-18]. This results in effective *p*-wave topological superconductors which support Majorana fermions.

Importantly, as shown by many experimental groups recently, topological insulators such as Bi$_2$Se$_3$ can become superconducting when doped by metals such as Cu, Sr and Nb[19-25]. Due to the non-trivial topology in the normal state band structures, the nature of pairing in these superconducting states has attracted much interest. In particular, it was proposed that Cu-doped Bi$_2$Se$_3$ could be a topological superconductor with odd parity pairing, which belongs to the A$_u$ representation of the D$_{3d}$ point group[26]. There are Majorana surface states associated with these superconducting phases that should cause zero-bias conductance peaks in tunneling experiments[21]. However, the zero-bias conductance peak found in an early experiment on $Cu_xBi_2Se_3$ is missing in more surface sensitive scanning tunneling experiments[27].

On the other hand, recent measurements on Cu-doped and Sr-doped $Bi_2Se_3$ observed that the in-plane three-fold rotational symmetry is spontaneously broken into a two-fold rotational symmetry below the superconducting transition temperature[28-31]. This spontaneous breaking of rotational symmetry can be explained if the superconducting phase is in the nematic phase, which belongs to the two-component $E_u$ representation of the point group[32, 33].

Interestingly, the $E_u$ representation, which is a two-dimensional representation, allows another topological superconducting phase, which spontaneously breaks the time-reversal symmetry. This phase is a nodal Weyl superconductor that supports Majorana arcs on the surfaces[33-37]. It was proposed by Qiu et al.[25] that Nb-doped $Bi_2Se_3$ spontaneously breaks time-reversal symmetry below $T_c$. The Nb-doped case may be special because of the finite magnetic moments of the Nb atoms intercalated in the van der Waals gap between the $Bi_2Se_3$ layers[25], which may enhance this Weyl superconducting phase[34]. On the other hand, other recent experiments on Nb-doped $Bi_2Se_3$, including torque magnetometry[24], penetration depth measurements[38, 39], and Andreev reflection spectroscopy[40], suggested that the system is in the nematic phase. However, only the torque experiments were directly probing the in-plane anisotropy, but the measurement were carried out in the irreversible regime and the results are difficult to interpret. In order to settle this issue we have measured the upper critical fields for magnetic fields applied in different directions in the basal plane, using magnetoresistance and DC magnetization measurements, and found a strong evidence of a two-fold rotational symmetry below $T_c$. The two-fold symmetry is further reflected in the DC magnetization value of the initial branch of the hysteresis loop after zero-field cooling, in the reversible thermodynamic contribution above the irreversibility field, as well as in the irreversible properties, such as the width of the hysteresis loop at a fixed magnetic field value. Our work thus provides further evidence that Nb-doped $Bi_2Se_3$ is in the odd-parity nematic superconducting phase. Due to the large superconducting volume fraction in Nb-doped $Bi_2Se_3$ compared to the Cu- and Sr-doped cases, Nb-doped $Bi_2Se_3$ can be an ideal material to investigate the topological properties of this nematic phase.

**RESULTS**

The electric resistivity of $Nb_{0.25}Bi_2Se_3$ (Sample 1) in zero magnetic field shows a superconducting transition in form of a sharp drop with onset at 3.2 K (see inset of Fig. 1). In Fig. 1 we show magnetoresistance data taken at fixed temperature of 0.34 K for different directions of in-plane magnetic fields varying over more than 180 degrees. A significant angular variation of the field-driven transition can be seen. The resistive increase occurs at ~0.4 T for orientations around 0°, and is shifted to higher fields until the field of the midpoint approximately doubles with a value exceeding 1.4 T peaking at ±90°. To define the upper critical field transition we used different criteria to determine characteristic fields from the data at which each angle a certain fixed percentage of the normal state resistance is reached: 50%, 75% and 90%. The so defined characteristic fields are presented in Fig. 2 as a function of the in-plane magnetic field orientation. The inset shows the same data in form of a polar plot. Centered at 0° a broad and rounded minimum is observed, while sharp peaks occur at ±90°. No signature of the three-fold crystalline symmetry is obvious. The data is dominated by a pronounced two-fold symmetry, which is completely in contradiction with the crystalline symmetry. Note that the normal state resistance above the upper critical field does not show any variation for the different in-plane orientations of the magnetic field, thus suggesting an isotropic normal state within the trigonal basal plane. The superconductivity thus appears to be nematic, and the in-plane angular dependence of the resistive transition is almost identical to what has been observed for $Sr_xBi_2Se_3$[30], although the upper critical field is clearly lower in $Nb_{0.25}Bi_2Se_3$.

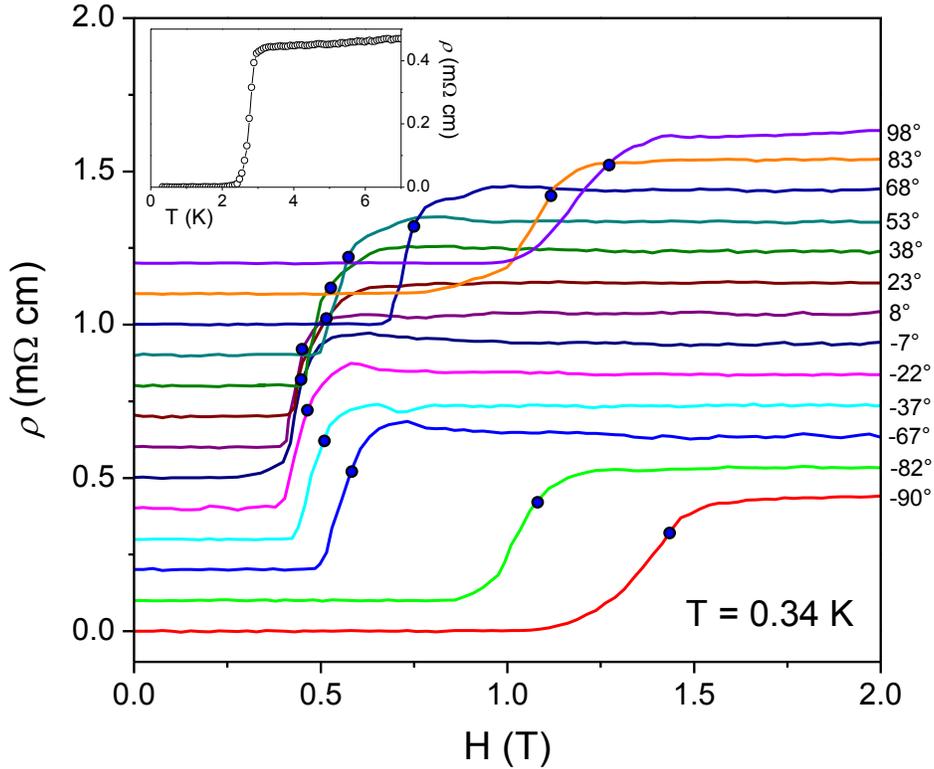

**Figure 1. Magnetoresistance of $Nb_{0.25}Bi_2Se_3$ as a function of in-plane magnetic field orientation.** All data were taken at $T = 0.34$ K. For clarity, offsets have been added to all data (except at -90°). The orientation of the magnetic field applied strictly parallel to the $Bi_2Se_3$ basal plane is marked on the right axis. The additional dots mark the points at which the magnetoresistance reaches 75% of the normal state resistance. Inset: zero-field resistivity showing the superconducting transition with onset at 3.2 K.

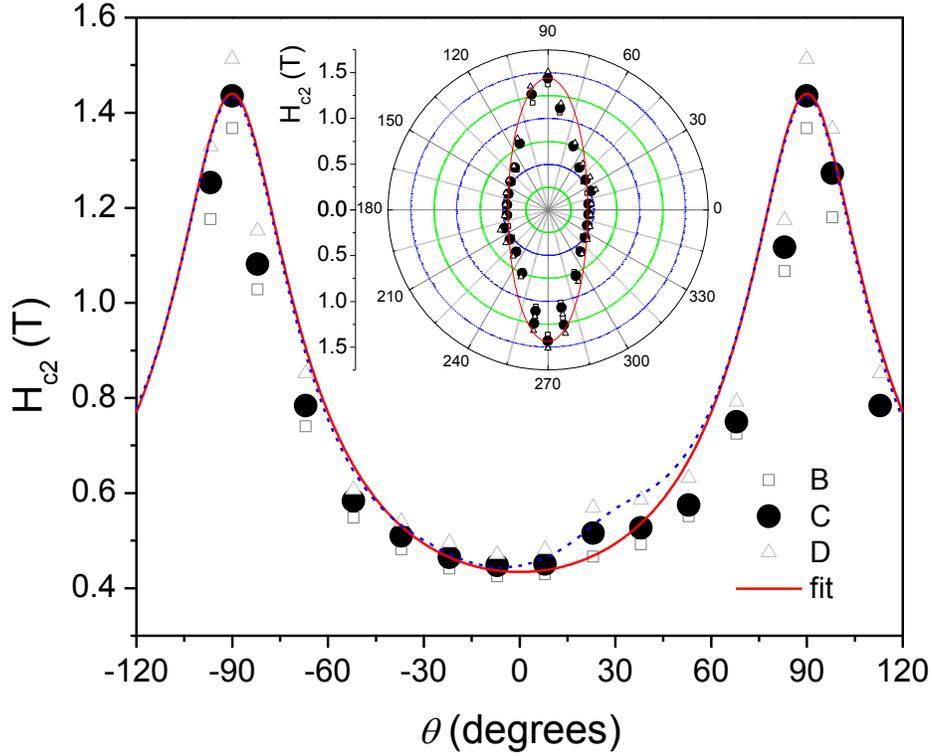

**Figure 2. Upper critical field as a function of the in-plane magnetic field orientation of Nb$_{0.25}$Bi$_2$Se$_3$.** The data was determined from the magnetoresistance data in Fig 1. The critical field was estimated according to various criteria when the magnetoresistance reaches 50% (squares), 75% (filled circles) and 90% (triangles) of the normal state value. The continuous line correspond to a theoretical model of nematic superconductivity with a single domain, while the dashed line considers an additional minority phase of ~10% volume fraction rotated by 60°, thus causing the additional bump at ~30° (see text for details). $\theta = 0$ is the normal direction to the mirror plane within the trigonal basal plane. The inset shows the same data in form of a polar plot.

In Fig. 3 we show a selection of magnetization hysteresis loop data of a second sample (Sample 2), which are taken at $T = 1.8$ K for different directions of the magnetic field applied parallel to the basal plane. The anisotropy of the upper critical field is evident. The data at the 101° angle have a clearly higher critical field, while the data at 0° show the lowest value as seen in the main figure showing an enlargement of the reversible regime of the magnetization above the irreversibility field $H_{irr}$ below which the hysteresis loops open. Data were collected under identical conditions at 0, 48, 101, 138, 180, 228, 281 and 318 degrees. For clarity, we have only included angles from -43° degrees (318° degrees) to 101° degrees, while larger angles follow exactly the same trend. The anisotropy is also reflected in the irreversible part, where the largest hysteresis occurs at 0° and the smallest at ~101°, as shown for the two extreme angles (0° and 101°) in the inset.

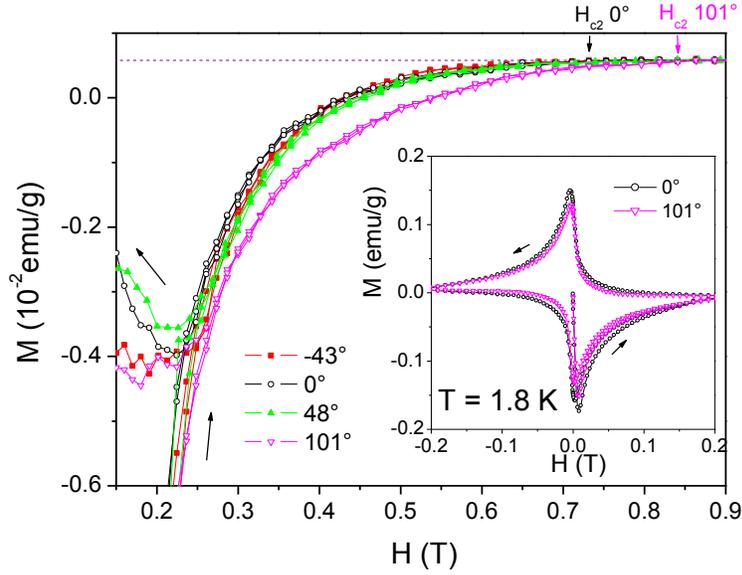

**Figure 3. Magnetization hysteresis loops of $Nb_{0.25}Bi_2Se_3$ for a selection of magnetic field directions in the basal plane.** The data were recorded at $T = 1.8$ K. The inset shows complete hysteresis loops for 0° and 101°, while the main graph shows an enlargement of the region in the range near $H_{c2}$ for -43° (317°), 0°, 48° and 101°. The dotted line marks the normal state limit.

In Fig. 4a we show the dependence of the upper critical field $H_{c2}$ on the orientation of the magnetic field vector in the basal plane over the full 360-degree range, and in Fig 4b the irreversibility fields $H_{irr}$ below which flux pinning becomes efficient, thus causing an opening of the hysteresis loops. Both data show a similar two-fold anisotropy. Upper critical field data of two different samples (Sample 2 & Sample 3) have been included in Fig 4a, which both show a very similar trend. In Fig. 4c, we show the magnetization value for Sample 2 from the initial zero-field-cooled branch of the magnetization loop at 8.5 mT, where the minimum associated with the lower critical field $H_{c1}$ occurs. The dependence on the in-plane direction of the magnetic field also reflects the two-fold anisotropy of the superconducting phase. The inset shows the corresponding magnetization data at 0° and 101°. The same field angular dependence is observed in the reversible part of the DC magnetization above the irreversibility field (Fig. 4d), which represents the purely thermodynamic contribution to the DC magnetization, and thus provides a bulk thermodynamic proof of nematic superconductivity.

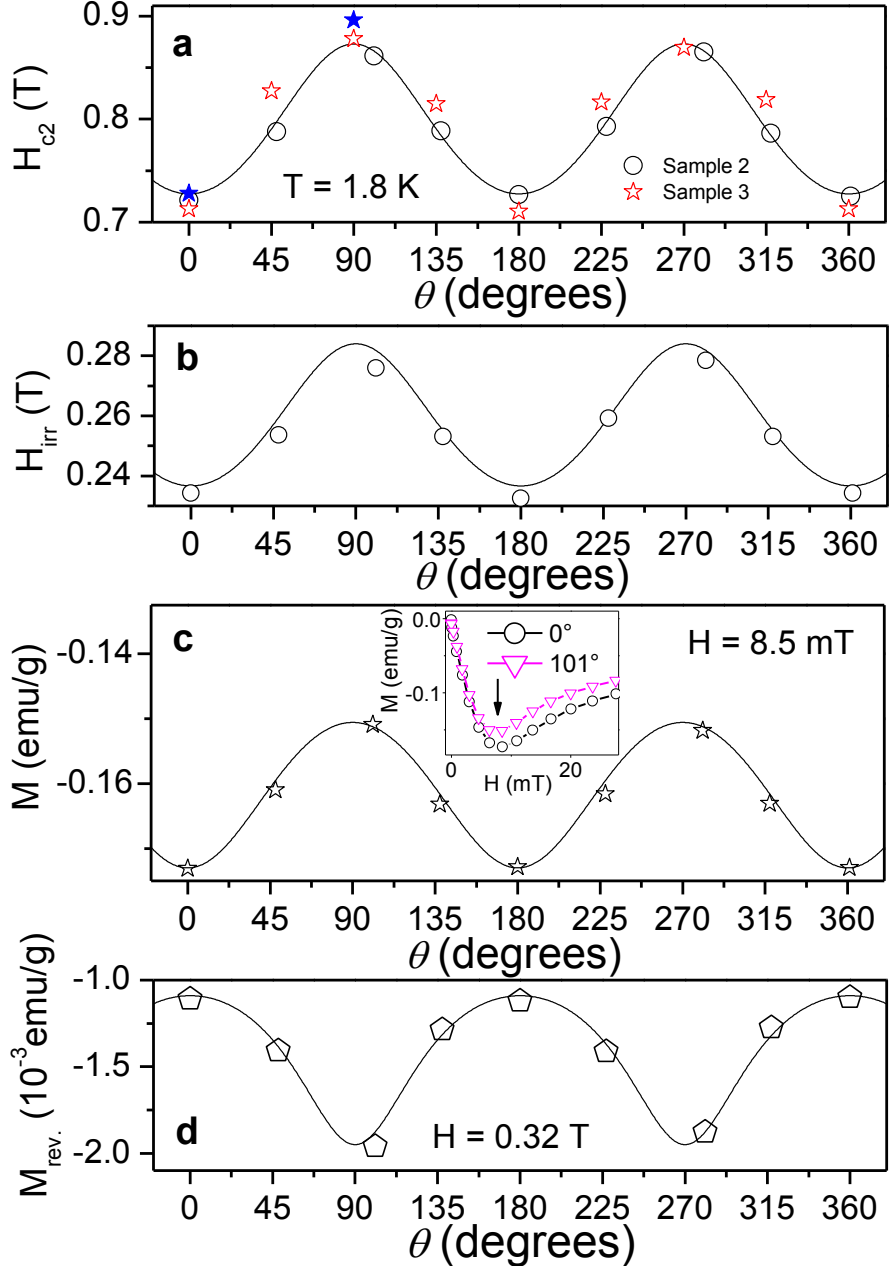

**Figure 4. Dependence of $H_{c2}$, $H_{irr}$ and the magnetization value on the in-plane magnetic field direction.** (**a**): $H_{c2}$ determined from the magnetization hysteresis loops at 1.8 K for Sample 2 (empty circles) and Sample 3 (empty stars). The two additional datapoints (filled stars) are taken from magnetoresistance data of Sample 3 as a check of consistency between the two methods. (**b**): Irreversibility fields $H_{irr}$ of Sample 2 below which the hysteresis loops open. (**c**): Dependence of the magnetization in the initial zero-field-cooled branch at 8.5 mT on the in-plane orientation of the magnetic field (Sample 2). The inset shows the corresponding magnetization data for 0° and 101°. (**c**): Dependence of the thermodynamic contribution of the magnetization taken in the reversible regime above $H_{irr}$ at 0.32 T on the in-plane orientation of the magnetic field (Sample 2). The lines in all graphs correspond to a theoretical model of nematic superconductivity (see text for details).

In Fig. 5. we show the temperature dependence of the upper critical field transition as obtained from the magnetoresistance for three different characteristic directions: for in-plane fields along -90° (Fig. 5a) and 0° (Fig. 5b), where the maximum and minimum of the critical field is observed, respectively, and for magnetic fields applied perpendicular to the $Nb_{0.25}Bi_2Se_3$ basal plane (Fig. 5c). The resulting phase diagram (Fig. 5d), compiled using the field at which the magnetoresistance reaches 90% of the normal state value, shows that the two-fold in-plane anisotropy is present at all temperatures below the superconducting transition temperature, although it becomes weaker with increasing temperature.

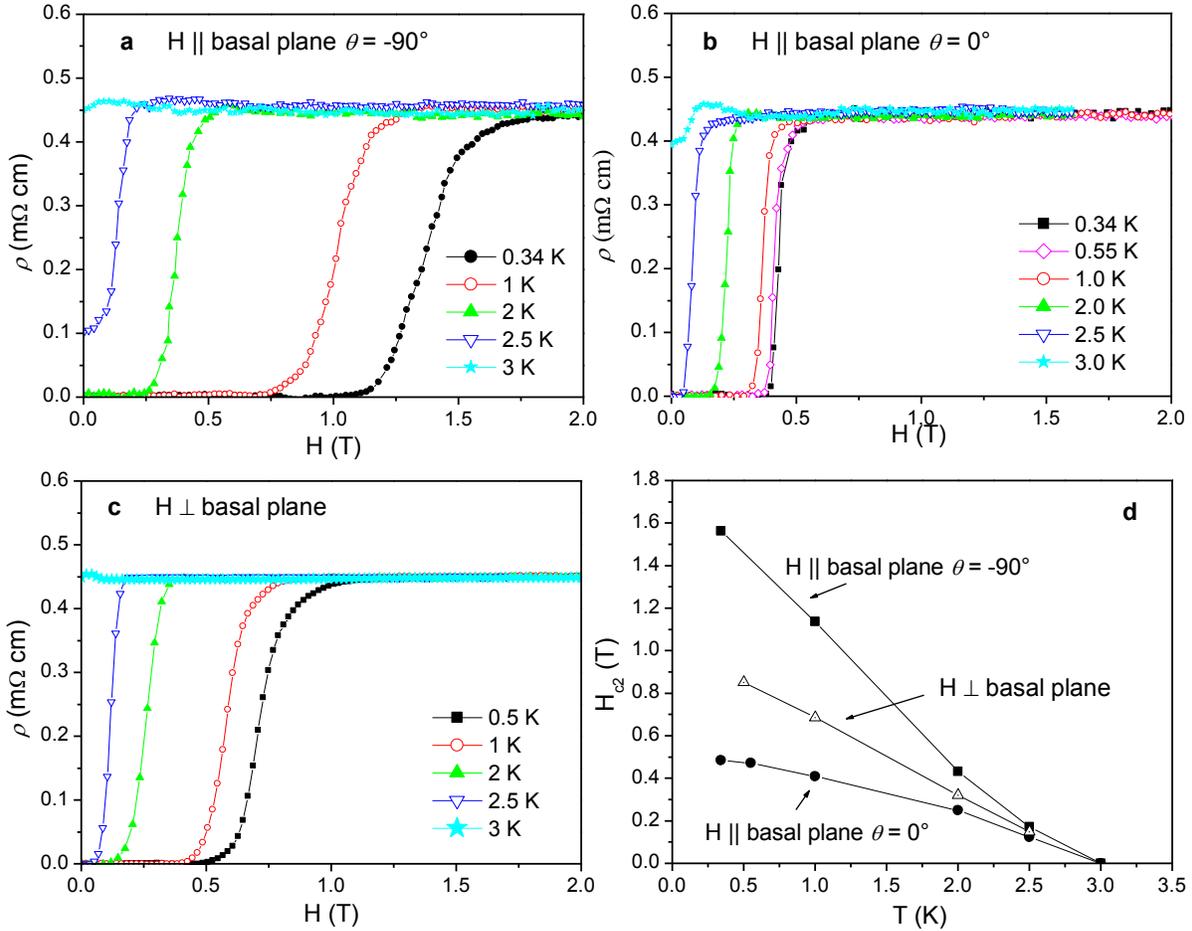

**Figure 5. Anisotropy of the temperature dependence of $H_{c2}$ of $Nb_{0.25}Bi_2Se_3$.** Magnetic field applied in-plane along -90° (**a**) and 0° (**b**) within the trigonal basal plane, and applied perpendicular to the $Bi_2Se_3$ basal plane (**c**). Panel **d** shows the corresponding magnetic field phase diagram. Here the criterion where the resistivity reaches 90% of the normal state resistance was chosen.

**DISCUSSION**

The field-angle dependence of the upper critical field of an unconventional superconductor directly reflects the order parameter symmetry[41,42]. Both, the magneto-resistive and the magnetic upper critical field transitions of $Nb_xBi_2Se_3$ show a clear two-fold symmetry instead of the expected three-fold symmetry, which suggests that the superconducting state of $Nb_xBi_2Se_3$ is nematic and spontaneously breaks the three-fold crystalline symmetry, in perfect agreement with what has been observed for $Cu_xBi_2Se_3$[28,29] and $Sr_xBi_2Se_3$[30,31]. Since the metal dopants are intercalated in the van der Waals gap between the $Bi_2Se_3$ layers, it has been argued that 1D clustering in the form of stripe like ion patters could be responsible for the two-fold anisotropy[30]. While we cannot rule this out from our experiments, the fact that the normal state resistance does not show any dependence on the in-plane field direction does make this scenario unlikely, and rather suggests a true nematic superconducting state. The superconducting Nb-doped $Bi_2Se_3$ thus appears to be very similar to $Cu_xBi_2Se_3$[28,29] and $Sr_xBi_2Se_3$[30,31], despite it has been shown that the Nb ions exhibit finite magnetic moments[25]. The latter can be seen in the form of a small paramagnetic contribution in the normal state magnetization in Fig. 6, which follows a Curie Weiss behavior as illustrated by the fit. The paramagnetic behavior is observed down to $T_c$ without any sign of magnetic ordering.

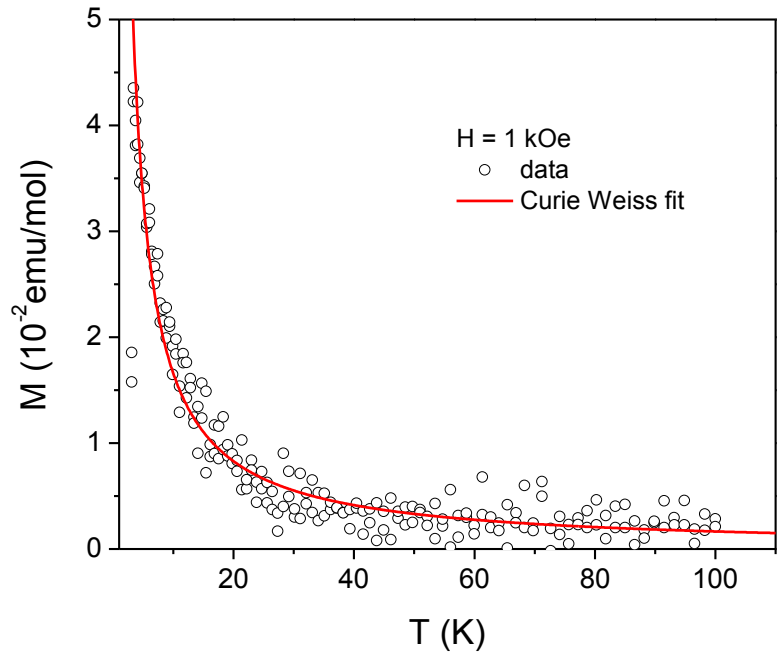

**Figure 6. Normal state DC magnetization.** The data were recorded in an applied field of 1 kOe on Sample 3. The data show a very weak paramagnetic contribution, as illustrated by the Curie Weiss fit.

The pronounced two-fold symmetry observed in in-plane upper critical fields can be explained consistently within the Ginzburg-Landau theory and could be identified as a signature of the nematic superconducting state. It is known that for a nematic superconductor at zero field, the two-component order parameter $(\eta_1, \eta_2)$ is pinned to $\eta_0(1,0)$ or $\eta_0(0,1)$ by sixth order terms in the Ginzburg-Landau free energy that arise from the crystalline anisotropy. Ref. 43 predicts that the in-plane upper critical field $H_{c2}$ of a nematic superconductor would have a six-fold symmetry, assuming that the nematic pinning effect on $(\eta_1, \eta_2)$ is negligible in the field-driven phase transition. This assumption, however, only applies if the superconducting phase is sufficiently close to the normal phase. The experimental data of $H_{c2}$ show a strong two-fold symmetry up to the 90% criterion, as shown in Fig. 2, indicating that to the precision of experiment, the pinning effect on the $(\eta_1, \eta_2)$ maintains. Based on these facts, we assume that $(\eta_1, \eta_2)$ is pinned to $\eta_0(1,0)$ or $\eta_0(0,1)$ and solve the corresponding Ginzburg-Landau equation. In this way one obtains the in-plane upper critical field as

$$H_{c2}(\Theta) = \frac{H_{c2}(0)}{\sqrt{\cos^2\Theta + \Gamma^2 \sin^2\Theta}}. \qquad (1)$$

Here $\Gamma$ indicates the in-plane anisotropy in the superconducting phase and the angle $\Theta$ is defined such that $\Theta = 0$ is the normal direction to the mirror plane. The details of the derivation from the Ginzburg-Landau theory can be found in the Methods section. The above formula well describes the two-fold symmetry in the in-plane upper critical fields of nematic superconductors. With this formula, we fit the experimentally measured in-plane upper critical field of Sample 1 as shown by the lines in Fig. 2, with fitting parameters: $\Gamma = 3.32$ and $H_{c2}(0) = 1.44$ T.

Of the three single crystals of different shapes we measured (see Methods section for details), no apparent correlation was observed with the orientation of the macroscopic crystal shape. In fact, the demagnetization factors for our samples with the parallel field orientation are hardly expected to have any significant influence on the in-plane variation of the magnetic properties. Indeed, the initial ZFC branch of the magnetization for the different in-plane field orientations all fall on top of each other (see inset of Fig. 4c). In any case, the upper critical field is not affected by the demagnetization factor and provides a solid proof of the nematic superconducting state. The theoretical model fits out data very well and further supports the existence of nematic superconductivity in Nb$_x$Bi$_2$Se$_3$. One open question is what determines the direction of the two-fold superconducting gap function within the three-fold symmetry of the basal plane. The gap function appears pinned along one of three identical crystalline directions and remains there during the entire experiment, even if the sample is brought into the normal state in between. Which direction the anisotropic superconducting gap function chooses likely only depends on microscopic details of the single crystal, such as surface roughness, defects, micro-cracks or local stress. However, the in-plane anisotropy for Sample 1 (64% at 1.8 K) is much greater than for Sample 2 and 3 (18 % and 21 % at 1.8 K, respectively). This is not an artifact from the different methods used since the upper critical field values for Sample 3 from magnetization and magnetoresistance agree very well (see Fig. 4a). This difference can only be understood if there are different domains in these samples in which the 2-fold gap function is pinned along different crystalline directions and partially cancel the macroscopic anisotropy. Sample 1 is probably almost a mono-domain sample with large anisotropy factor $\Gamma = 11$, which causes the particularly sharp maxima in Fig. 2. Otherwise, additional smaller peaks should occur at ±30°. In fact, there is actually a small bump at +30 degrees, which could indicate a ~10% volume fraction of a minority domain rotated by 60° with respect to the majority domain. This is illustrated by the dashed line in Fig. 2, which represents a fit that takes into account such an admixture of a minority phase. However, this feature is very weak and at the resolution limit. For samples 2 and 3, the maxima are much broader, which is taken into account in the fitting curve in Fig. 4a by considering a smaller anisotropy factor $\Gamma = 1.44$. For a small $\Gamma$ value, the cosine term dominates the angular in-plane $H_{c2}$ dependence. In the presence of minority domains, the individual sinusoidal terms would basically merge together without causing any side maxima in the angular $H_{c2}$ dependence, but reducing the overall variation. What causes the strong variation in $\Gamma$ is unclear, but Sample 1 is likely to be of higher crystalline quality, as evidenced by its shiny flat surfaces. Sample 2 and 3 have rather rough surfaces, and surface roughness, along with internal crystal irregularities, could lead to a broadening of the 2-fold gap structure, in addition to the occurrence of minority domains.

To summarize, we have carefully determined the field-angular dependence of the magnetoresistive and the magnetic upper critical field transitions of Nb-doped $Bi_2Se_3$ with 0.25 Nb atoms per formula unit. The in-plane angular dependence when the field is applied strictly parallel to the $Nb_{0.25}Bi_2Se_3$ basal plane shows a pronounced two-fold symmetry very similar to $Cu_xBi_2Se_3$[28, 29] and $Sr_xBi_2Se_3$[30, 31], and thus provides further experimental evidence from two different experimental methods that nematic superconductivity also exists in $Nb_xBi_2Se_3$. The two-fold symmetry is also reflected in the absolute value of the magnetization (for example in the initial curve of the hysteresis after zero-field cooling, as well as in the thermodynamic reversible regime above the irreversibility field) and in the anisotropic characteristics, such as the width of the hysteresis loop and the irreversibility field. The in-plane anisotropy of the upper critical field can be perfectly fitted with a theoretical model[32, 33, 43] for nematic superconductivity. The existence of magnetic moments without macroscopic magnetic order could provide a way of tuning the superconducting properties, e.g. by varying the Nb concentration or by introducing different magnetic ions with stronger moments to see whether the unconventional pairing symmetry of the superconducting state could be dramatically altered by the internal magnetic fields.

**METHODS**

The detailed growth method and characterization of $Nb_{0.25}Bi_2Se_3$ in the single crystalline form can be found in Refs. 24, 25. Measurements have been done on three different samples of dimensions and geometry as illustrated in Fig.7. The magnetic field directions of 0 and 90 degrees are marked by arrows.

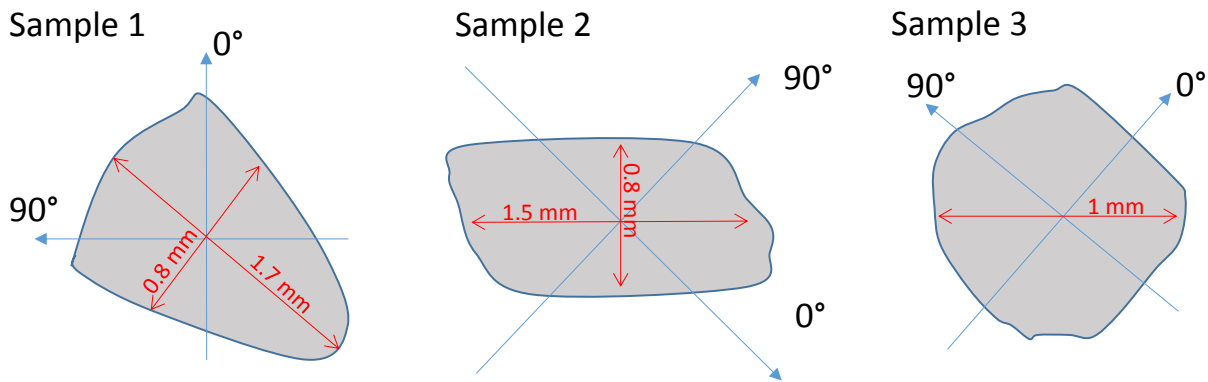

**Figure 7. Illustration of the sample geometries and shapes used in this study.** The long arrows indicate the in-plane magnetic field orientations. Short arrows mark the approximate in-plane dimensions. The sample thickness were 0.2 mm (Sample 1), 0.1 mm (Sample 2) and 0.3 mm (Sample 3).

**Electrical transport**

We carried out standard four-probe resistance measurements in a 15 T magnet cryostat with $^3$He variable temperature insert. The single crystal (Sample 1) of approximate dimensions 1.7 x 0.8 x 0.2 mm$^3$ and demagnetization factor of ~0.7 was mounted on an Attocube ANR51/RES precise nano rotary stepper with resistive encoder, which provides a millidegree precision with integrated angle measurement. The current was injected along the long direction (Fig. 7) and coincided with the magnetic field direction for an in-plane field orientation of 225°. The Nb$_{0.25}$Bi$_2$Se$_3$ basal plane was aligned precisely parallel to the magnetic field and the rotator allowed us to rotate the sample so that the magnetic field direction could be varied within the basal plane. In a subsequent separate measurement, we oriented the same device with the basal plane perpendicular to the field to derive the complete temperature dependence of upper critical fields along different crystalline directions. The measurements were performed with both an AC and a DC method to check for consistency. All resistance data shown in this article were measured with the AC technique with an alternating current of 0.1 mA amplitude and a frequency of 17.7 Hz using a Keithley 6221 AC/DC current source. The signal was sent through a bandpass filter and measured by a SRS830 digital lock-in amplifier.

**DC magnetization**

The DC magnetization was measured on two single-crystalline samples of 847 μg (Sample 2) and 4.5 mg (Sample 3) mass with a Quantum Design Vibrating Sample SQUID magnetometer (VSM-SQUID). The measurement was deliberately performed on samples of different shapes and size to eliminate any effects of the sample shape and quality. The shape of Sample 2 can be approximately described as a rectangular prism of 1.5 x 0.8 x 0.1 mm$^3$ with demagnetization factor of ~0.8. Sample 3 yields an approximately cylindrical shape of ~ 1 mm diameter and 0.3 mm height with demagnetization factor of ~0.6. The samples were first cleaved to obtain a shiny and very flat bottom plane of the crystal, which was then fixed with vacuum grease to the flat surface of a quartz sample holder so that the basal plane of Nb$_{0.25}$Bi$_2$Se$_3$ was oriented parallel to the applied magnetic field in all experiments. The vacuum grease allowed us to rotate the sample at room temperature by carefully pushing the sample with a toothpick into the desired angular orientation characterized by the in-plane angle $\theta$, carefully avoiding a contact with the grease. The sample holder was then inserted into the magnetometer. The sample was cooled to 1.8 K in zero field and a full hysteresis loop with maximum field of ±1 T was measured. Subsequently, the sample was heated above $T_c$ and again cooled in zero field to repeat the measurement, but with reverse field direction to obtain data at an angle of $\theta$ + 180 degrees. After this measurement cycle, the sample holder was removed, the sample was rotated by ~45 degrees, and the procedure was repeated until the full 360 degrees angle range was obtained. The exact angular orientations were determined with an accuracy of 0.5 degree from a photograph by measuring the angle between a linear edge of the sample with respect to the sample holder using a protractor. The absolute values of the in-plane field direction $\theta$ are chosen so that $\theta = 0$ is the normal direction to the mirror plane within the trigonal basal plane.

**Theory**

In the Ginzburg-Landau theory for a superconductor, the phenomenological free energy density $f_{tot}$ is the sum of a homogeneous term $f_{hom}$ and a gradient term $f_D$ so that $f_{tot}= f_{hom}+ f_D$. The superconducting $Nb_xBi_2Se_3$ has a crystal structure belonging to the $D_{3d}$ point group. In the $E_u$ representation, the homogeneous part of the free energy density up to the sixth order has the form[33]

$$f_{hom} = A\left(|\eta_1|^2 + |\eta_2|^2\right) + B_1\left(|\eta_1|^2 + |\eta_2|^2\right)^2 + B_2\left|\eta_1^*\eta_2 - \eta_1\eta_2^*\right|^2$$
$$+C_1\left[\left(\eta_+^*\eta_-\right)^3 + \left(\eta_-^*\eta_+\right)^3\right] + C_2\left(|\eta_1|^2 + |\eta_2|^2\right)^3 + C_3\left(|\eta_1|^2 + |\eta_2|^2\right)\left|\eta_1^*\eta_2 - \eta_1\eta_2^*\right|^2 \quad (2)$$

where $A$, $B_{1,2}$ and $C_{1,2,3}$ are the Ginzburg-Landau coefficients and $\eta_\pm = \eta_1 \pm i\eta_2$. Here $A \propto (T-T_c)$, $B_2>0$ for the nematic phase, and $C_1$ is responsible for pinning the nematic state from $(\eta_1,\eta_2)$ to $\eta_0(1,0)$ or $\eta_0(0,1)$. Defining the covariant derivative $D_i = -i\partial_i - qA_i$, where $A_i$ is the electromagnetic vector potential and $q=2e$, the gradient term, $f_D$ can be written as[43]

$$f_D = J_1(D_i\eta_a)^* D_i\eta_a + J_2\varepsilon_{ij}\varepsilon_{ab}(D_i\eta_a)^* D_j\eta_b + J_3(D_z\eta_a)^* D_z\eta_a + J_4\left(\tau_{ij}^z\tau_{ab}^z + \tau_{ij}^x\tau_{ab}^x\right)(D_i\eta_a)^* D_j\eta_b$$
$$+J_5\left[\tau_{ab}^x(D_x\eta_a)^* D_z\eta_b + \tau_{ab}^x(D_z\eta_b)^* D_x\eta_a + \tau_{ab}^z(D_y\eta_a)^* D_z\eta_b + \tau_{ab}^z(D_z\eta_b)^* D_y\eta_a\right] \quad (3)$$

where we set $i = x, y$, and $a = 1, 2$, and $J_1$, $J_2$, $J_3$, $J_4$, $J_5$ are the phenomenological Ginzburg-Landau coefficients. Here $\varepsilon_{(i,j),(a,b)}$ is the anti-symmetric tensor and $\tau_{(ij),(ab)}^{(x),(z)}$ is the Pauli matrix acting on the index $(ij)$ and $(ab)$. In the presence of the in-plane magnetic field $\vec{H} = H(\cos\Theta, \sin\Theta, 0)$ with the gauge $\vec{A} = Hz(\sin\Theta, -\cos\Theta, 0)$, the covariant derivative is defined as $D_x = -i\partial_x + 2eHz\sin\Theta$, $D_y = -i\partial_y - 2eHz\cos\Theta$ and $D_z = -i\partial_z$. The linearized Ginzburg-Landau equation can be derived from the functional variation with respect to the pairing field $\frac{\delta f_{tot}}{\delta \eta_a^*} = 0$ so that

$$-A\vec{\eta} = \left[\left(J_1 D_\perp^2 + J_3 D_z^2\right)\tau^0 - J_4 D_\perp^2\left(\cos 2\Theta\,\tau^z + \sin 2\Theta\,\tau^x\right) + J_5\{D_z, D_\perp\}\left(-\cos\Theta\,\tau^z + \sin\Theta\,\tau^x\right)\right]\vec{\eta}, \quad (4)$$

where $\vec{\eta} = (\eta_1, \eta_2)^T$ and $D_\perp = D_x\sin\Theta - D_y\cos\Theta$. The parallel covariant derivative terms $(D_x\sin\Theta - D_y\cos\Theta)\eta_a$ are dropped. We set $(\eta_1, \eta_2)$ to be $\eta_0(0,1)$ (which is a fully gapped state[31, 32]) and the coupled Ginzburg-Landau equations reduce to one equation

$$-A\eta_0 = \left(J_1 D_\perp^2 + J_3 D_z^2 + J_4 D_\perp^2 \cos 2\Theta + J_5\{D_z, D_\perp\}\cos\Theta\right)\eta_0. \quad (5)$$

The right-hand side of the equation above can be treated as the Hamiltonian for a simple Harmonic oscillator, and the in-plane upper critical field can be obtained from its ground state energy level

$$H_{c2}(\Theta) = \frac{H_{c2}(0)}{\sqrt{\cos^2\Theta + \Gamma^2 \sin^2\Theta}} \ . \tag{6}$$

Here the dimensionless parameter $\Gamma^2 = \frac{J_1 J_3 - J_3 J_4}{J_1 J_3 + J_3 J_4 - J_5^2}$ indicates the in-plane anisotropy and $H_{c2}(0) = -\frac{A}{2e\sqrt{J_1 J_3 + J_3 J_4 - J_5^2}}$. The in-plane upper critical field has the same form for the nematic phase $\eta_0(1,0)$ while the definitions of related parameters are different.

## ACKNOWLEDGMENTS

R.L. thanks U. Lampe for technical support. This work was supported by grants from the Research Grants Council of the Hong Kong Special Administrative Region, China (SBI15SC10, HKUST3/CRF/13G, C6026-16W). Y.S.H. acknowledges support from National Science Foundation DMR-1255607. K.T.L. acknowledges the support of the Croucher Foundation and the Dr. Tai-Chin Lo Foundation.

## AUTHOR CONTRIBUTIONS

J.S. performed the electrical transport experiments with help of C.-W.C. R.L. performed the magnetization measurements. S.H. Lee and Y.S.H. did the sample preparation and characterization. J.S., R.L., C.-W.C and H.Z. analyzed the data. W.Y.H., N.F.Q.Y. and K.T.L. did the theoretical interpretation and fitting of the data. R.L. and K.T.L. contributed to the original idea and supervised the project. J.S., R.L., W.Y.H. and N.F.Q.Y. wrote the manuscript. All authors contributed to the discussion and data interpretation and have read and approved the final manuscript.